\documentclass[aps,prb,preprint,showpacs,groupedaddress,floatfix]{revtex4}

\usepackage{amsmath}
\usepackage{amssymb}
\usepackage{graphicx}
\usepackage{dcolumn}
\usepackage{bm}
\usepackage{subfigure}

\begin{document}

\title{Constraints on the phase diagram of molybdenum from \\
       first-principles free-energy calculations}

\author{C. Cazorla$^{1,2}$}
\author{D. Alf\`e$^{1,2,3,4}$}
\author{M. J. Gillan$^{2,3,4}$}
\affiliation{
$^{1}$Department of Earth Sciences, UCL, London, WC1E 6BT, UK\\
$^{2}$Thomas Young Centre at UCL, London WC1E 6BT, UK \\
$^{3}$London Centre for Nanotechnology, UCL, London WC1H 0AH, UK \\
$^{4}$Department of Physics and Astronomy, UCL, London WC1E 6BT, UK}
\pacs{64.10.+h,64.70.D-,64.70.K-,71.15.Pd}

\begin{abstract}

We use first-principles techniques to re-examine the suggestion
that transitions seen in high-$P$ experiments on Mo are solid-solid
transitions from the bcc structure to either the fcc or hcp structures.
We confirm that in the harmonic approximation the free energies
of fcc and hcp structures become lower than that of bcc at
$P > 325$~GPa and $T$ below the melting curve, as reported recently.
However, we show that if anharmonic effects are fully included
this is no longer true. We calculate fully anharmonic free energies
of high-$T$ crystal phases by integration of the thermal average
stress with respect to strain as structures are deformed into
each other, and also by thermodynamic integration from harmonic
reference systems to the fully anharmonic system. Our finding
that fcc is thermodynamically less stable than bcc in the relevant
high-$P$/high-$T$ region is supported by comparing the melting curves
of the two structures calculated using the first-principles
reference-coexistence technique. We present first-principles
simulations based on the recently proposed Z method which also support
the stability of bcc over fcc.

\end{abstract}

\maketitle

\section{Introduction}
\label{sec:introduction}

\begin{figure}
\centerline{
\includegraphics[width=0.8\linewidth]{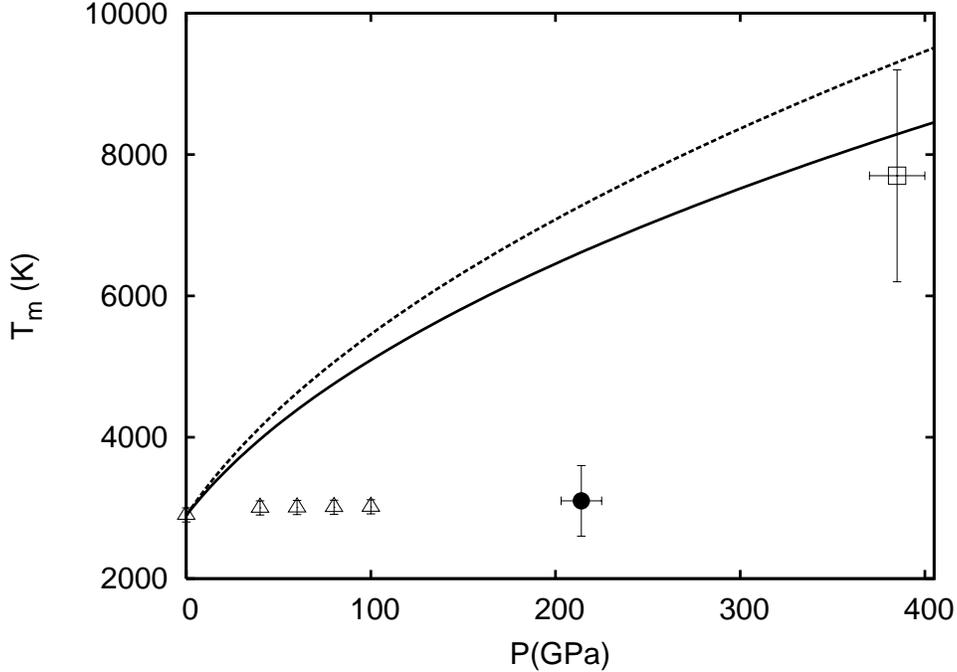}}%
\caption{Points on the solid-liquid boundary of Mo as observed in 
DAC ($\triangle$~[\onlinecite{errandonea01}]) and shock wave  
($\square$~[\onlinecite{hixson92}]) experiments.
The shock wave datum ($\bullet$~[\onlinecite{hixson92}]) obtained at low-$P$ 
has been interpreted as a solid-solid phase transition.    
Lines represent theoretical predictions of the melting curve of bcc
Mo by Cazorla \emph{et al.} (solid line~[\onlinecite{cazorla07}])
and Belonoshko \emph{et al.} (dashed line~[\onlinecite{belonoshko08}]).}
\label{figintro}
\end{figure}

The past $10$~years have seen a lively controversy over the phase
diagrams of transition metals at megabar pressures, with the
pressure dependence of the melting temperature $d T_{\rm m} / d P$
from diamond-anvil-cell (DAC) measurements differing greatly from that deduced 
from shock data and from first-principles calculations 
(see Fig.~\ref{figintro}).~\cite{errandonea01,alfe02b,burakovsky10,dewaele10,wu09,cazorla07,taioli07,errandonea05,santamaria09,belonoshko08,liu08,duffy05}
One suggested resolution of the controversy is that the transition
interpreted as melting in some of the DAC experiments may in fact
be a solid-solid crystallographic transformation, and in the
case of molybdenum this is consistent
with the observation of two transitions in 
shock experiments.~\cite{hixson92,mitchell81}
This suggestion appeared to be confirmed by recent first-principles
work on Mo, which indicated a transition from the low-temperature
body-centred-cubic (bcc) structure to a close-packed structure in the
appropriate temperature region.~\cite{belonoshko08,cazorla08b,zeng10} 
However, that work relied on two important assumptions, which we examine in 
detail in this paper. 
The results we shall present imply that those assumptions and the
conclusions drawn from them may be incorrect, so that further work
is still needed to resolve the controversy.

DAC measurements have been reported on the melting curves of several
transition metals, including Ti, V, Cr, Fe, Co, Ni, Mo, Ta and 
W.~\cite{errandonea01,santamaria09,japel05,errandonea03} 
The measurements extend up to nearly $100$~GPa ($1$~Mbar), and in most cases
the increase of melting temperature $T_{\rm m}$ between ambient
pressure and $100$~GPa is surprisingly small; in the case of Mo, the
increase is only $\sim 200$~K.~\cite{errandonea01,santamaria09} 
These findings are in stark contrast to the melting curves deduced from shock 
measurements, which are
available for Fe, Mo, Ta and W.~\cite{hixson92,mitchell81} 
For Mo, the increase of $T_{\rm m}$
between ambient and $100$~GPa estimated from shock data is $\sim 2000$~K.
Recently, density functional theory (DFT) calculations of the melting curves of Mo and Ta have been 
reported.~\cite{belonoshko04,cazorla07,taioli07,belonoshko08,burakovsky10} 
The DFT predictions are expected to be reliable, because
it is well known that DFT, without any adjustable parameters,
gives excellent results for a wide range of properties of transition
metals, including cold compression curves up to 
$\sim 300$~GPa,~\cite{cazorla07,taioli07,stixrude94,vocadlo00}
Hugoniot $P ( V )$ curves,~\cite{taioli08,cazorla08c,alfe02b}  phonon dispersion relations (and their
pressure dependence, in the case of Fe),~\cite{alfe00,cazorla07,taioli07,vocadlo08,mao01} and low-temperature
phase boundaries.~\cite{cazorla08} Furthermore, techniques for calculating melting
curves using DFT have become firmly established over the past $10$ years 
and more, and are known to give accurate 
results.~\cite{sugino95,dewijs98,alfe99,dewaele07,vocadlo02,vocadlo04,gillan06}
The DFT results for $T_{\rm m} ( P )$ of Mo and Ta lend
support to the correctness of the melting curves deduced from 
shock data.~\cite{cazorla07,taioli07,belonoshko08,burakovsky10}


It has become clear very recently that experimental difficulties may have
led to a substantial underestimate of high-$P$ melting temperatures
in earlier DAC measurements. The work of Dewaele {\em et al.}~\cite{dewaele10}
indicates that formation of metal carbide by chemical reaction 
between the diamond
and the metal sample can be a major problem. Their work also shows
that difficulties in the pyrometric measurement of temperature
can also lead to substantial underestimates of $T_m$. In the case of Ta,
their measurements give a melting curve that is well above those given
by earlier DAC experiments and is fairly close to (though systematically
lower than) the predictions from DFT. Nevertheless, in the case of Mo,
the occurrence of two breaks in the shock data seems to leave little
doubt that there is a transition from the low-$T$ bcc structure
to an unidentified high-$T$ crystal structure, followed by the melting
of the latter. The $T$ and $P$ of the lower transition (3500~K and
200~GPa) lie close to the natural extrapolation of the $T ( P )$
boundary identified as melting in the older DAC experiments. This suggests
that this boundary is associated with a bcc-solid transition. Even in
the case of Ta, recent evidence based on DFT 
calculations~\cite{burakovsky10} indicates
that a bcc-solid transition is implicated in earlier DAC attempts
to detect high-$P$ melting.

To substantiate the picture of a bcc-solid transition followed
by a melting transition, it is necessary to show that 
another crystal structure
becomes thermodynamically more stable than bcc at high temperatures,
and to identify this structure. This was the aim of the recent DFT
work on Mo by Belonoshko {\em et al}.~\cite{belonoshko08} They 
showed that, in the quasiharmonic
approximation, the Gibbs free energy of the fcc structure is lower than
that of bcc over a substantial high-$P$/high-$T$ region of the phase 
diagram below the bcc melting curve. The fcc structure becomes
harmonically unstable (there are imaginary phonon frequencies) for
$P < 350$~GPa, but extrapolation of the predicted bcc-fcc phase boundary
passes quite close to the $( P , T )$ of the lower shock transition.
The quasiharmonic calculations on the bcc and fcc free energies were
independently confirmed by two subsequent 
papers.~\cite{cazorla08b,zeng10}
As independent evidence that fcc is more stable than bcc at high~$T$, 
Belonoshko {\em et al.}~\cite{belonoshko08}
used the Z method~\cite{belonoshko06,davis08} to calculate the melting 
curve of fcc Mo. (The Z method employs observations of spontaneous
melting of the superheated solid in constant-energy molecular
dynamics simulations to
determine points on the melting curve.) They found that the fcc melting
curve lies above the bcc melting curve, thus appearing to confirm
that the free energy of fcc is lower than bcc. However, we note the 
two important assumptions made here:
first, that anharmonic contributions to the free energies of high-$T$ bcc
and fcc Mo can be neglected; and second, that the 
first-principles statistical-mechanical techniques that were employed
have the precision needed to distinguish between the 
possibly rather similar melting
curves of bcc and other stuctures. There is also the
question of whether fcc can remain vibrationally stable at high~$T$
in the region $P < 350$~GPa, where the harmonic 
phonons are unstable. These are the
issues addressed in the present paper.

We use two methods here for comparing the free energies of the bcc and
other crystal structures, and both methods fully include anharmonicity.
The other structures we examine are fcc and hexagonal-close-packed (hcp).
The first method uses the fact that the free 
energy difference between two systems that differ only by a finite
strain can be obtained by integrating the thermal average stress with
respect to strain.~\cite{nielsen85,alfe98} We use this idea to obtain the free energy difference
between fcc and bcc, which can be transformed into each other by a continuous
strain along the Bain path (BP).
The second method employs thermodynamic integration (TI) from a harmonic
reference system to the fully anharmonic system described by
DFT. This is essentially the same method as we employed in earlier work
on Fe.~\cite{alfe99,vocadlo03,alfe02b} As further ways of probing the possible
thermodynamic stability of the fcc structure, we have re-examined the
melting curve of fcc Mo, using the reference coexistence technique
employed in some of our earlier work,~\cite{gillan06,vocadlo04,alfe04,alfe02a} 
and we have also performed our own first-principles Z-method calculations 
on the melting of bcc and fcc Mo.
All the results point to the conclusion that none of the other structures 
is thermodynamically more stable than bcc at high~$T$. 

The rest of the paper is organised as follows. In 
Sec.~\ref{sec:harmonic}, we summarise
briefly the technical details of the DFT methods employed in all the
calculations, and we then present our results for the harmonic dispersion
relations of the bcc, fcc and hcp structures over a wide range
of $P$; we note the pressure thresholds below which the fcc and hcp
structures become harmonically unstable. In the same Section, we
report our results for the harmonic free energies, and hence the
predicted phase boundaries separating the different structures.
Sec.~\ref{sec:stability} presents our results on the 
vibrational, elastic and thermodynamic
stability of the different structures, including our Bain-path calculations
of the free energy differences between fcc and bcc. Our calculations
of the anharmonic contributions to the free energy, and the effect
of these contributions on the phase boundaries 
are presented in Sec.~\ref{sec:anharmonic}.
Our reference coexistence calculations of the fcc melting curve and 
the comparison with the bcc melting curve obtained already by the same 
technique are outlined in Sec.~\ref{sec:melting}, where we also
present our new Z-method calculations. 
Finally, we draw all the results 
together in Sec.~\ref{sec:discussion}, and suggest 
what future investigations might help 
to resolve the controversies over the phase diagram of Mo and 
other transition metals.

\section{Harmonic calculations}
\label{sec:harmonic}

\subsection{DFT techniques}
\label{subsec:dfttechniques}
 
All calculations were done 
using the projector augmented wave version of DFT as 
implemented in the VASP package.~\cite{blochl94,kresse96}
All atomic states up to and including $4s$ were treated as
core states, with $4p$ and all higher states being valence states.
We used the PBE form of generalised gradient approximation
to the exchange-correlation functional.~\cite{perdew96}
An energy cut-off of $224.6$~eV was used throughout; the 
adequacy of this value was shown in a previous 
work where we performed extensive numerical convergence 
tests.~\cite{cazorla07}
Dense Monkhorst-Pack grids~\cite{monkhorst76} 
were used for electronic $k$-point sampling in static
perfect lattice calculations, to guarantee convergence of 
the total energy to better than $1$~meV/atom. 
Thermal excitation of electrons was included via the finite-$T$
version of DFT originally developed by Mermin.~\cite{mermin65,alfe01}   
Phonon frequencies in our calculations were obtained by
the small-displacement method~\cite{kresse95,phon} using large supercells. 
For molecular dynamics (MD) simulations, we used the Born-Oppenheimer 
scheme where the self-consistent ground state is recalculated at each
MD time-step. These simulations were performed in the microcanonical
$(N, V, E)$ and canonical $(N, V, T)$ ensembles; temperatures in 
$(N, V, T)$ simulations were maintained using Nos\'{e} thermostats. 
Values of the technical parameters (duration of the MD runs,
{\bf k}-point grids, number of atoms, etc.) will be presented 
together with the results.

\subsection{Harmonic free energies and phase boundaries}
\label{subsec:harmonic_free_energy}

\begin{figure}
\centerline{
\includegraphics[width=0.8\linewidth]{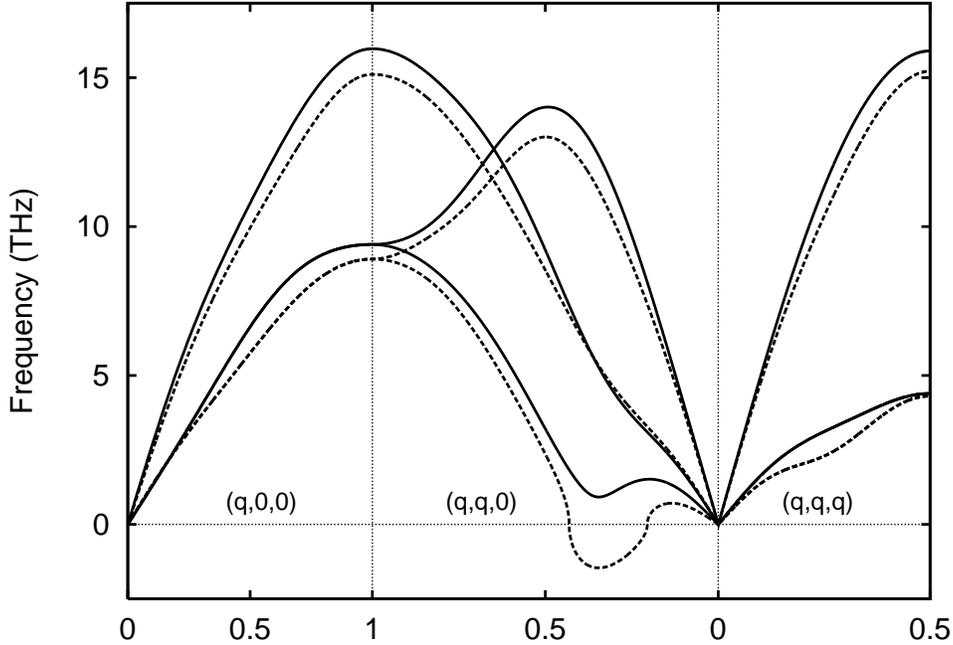}}%
\caption{\emph{Ab initio} vibrational phonon frecuencies of Mo in the fcc 
         structure calculated at volumes $V = 9.64$~\AA$^{3}$/atom 
         ($P = 328$~GPa, solid line) and $V = 10.19$~\AA$^{3}$/atom 
         ($P = 265$~GPa, dashed line).}
\label{fig1}
\end{figure}

\begin{figure}
\centerline{
\includegraphics[width=0.8\linewidth]{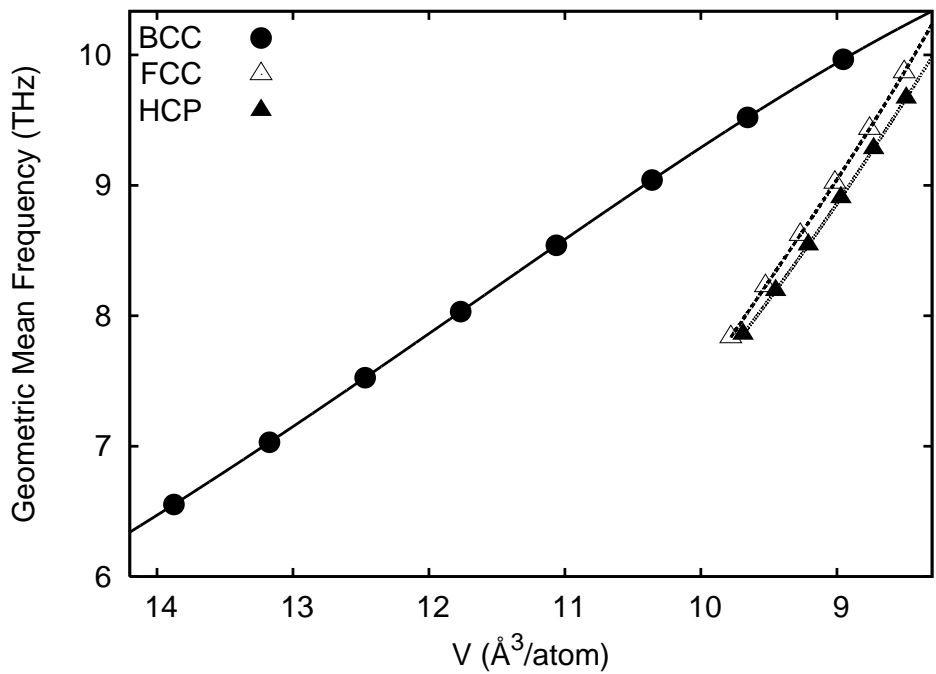}}%
\caption{Geometric mean frequency $\bar{\omega}$ of Mo in the 
         bcc, fcc and hcp structures as a function of volume.
         Symbols represent states at which the calculations
         have been carried out and the lines are guides to the
	 eye.}
\label{fig2} 
\end{figure}

\begin{figure}
\centerline{
\includegraphics[width=0.8\linewidth]{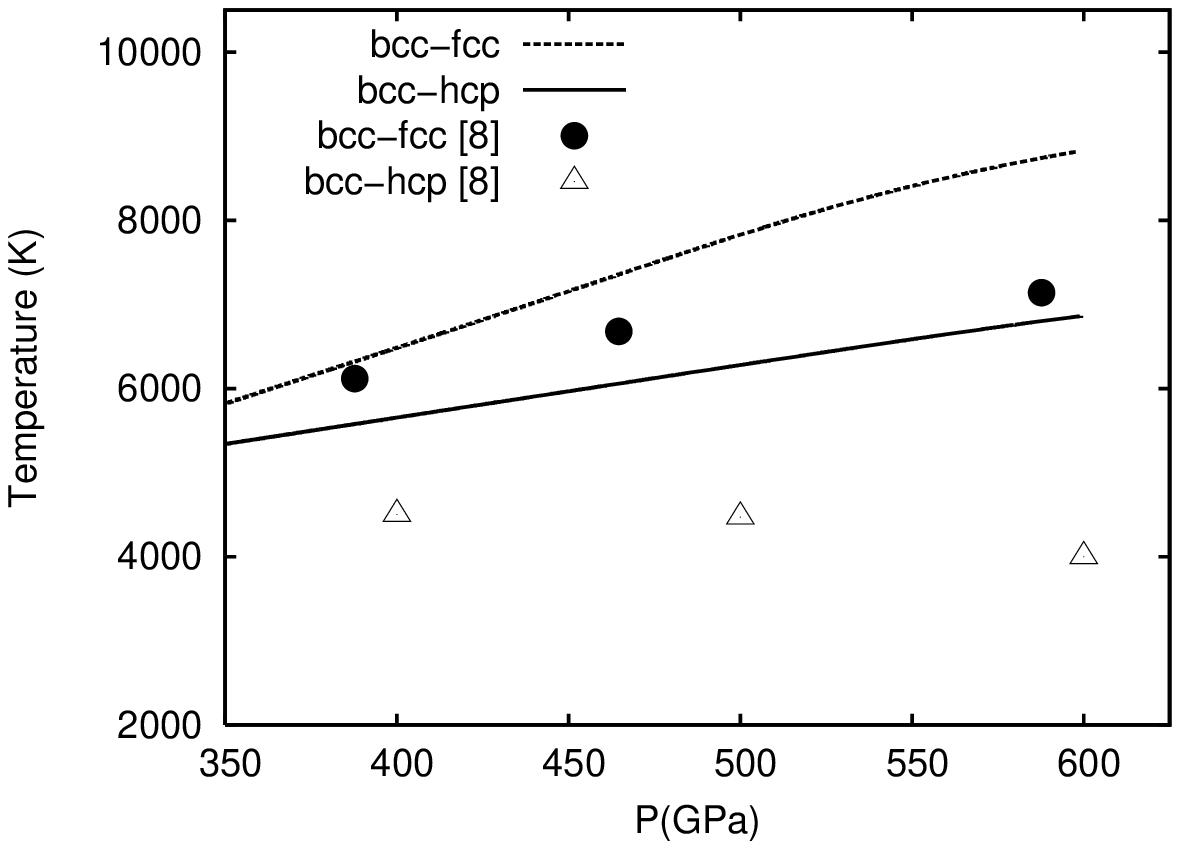}}%
\caption{Solid-solid phase boundaries in Mo at high~$P$ and high~$T$
         as obtained with first-principles harmonic free-energy calculations.
         Results obtained by Belonoshko \emph{et al.}~[\onlinecite{belonoshko08}]
         are shown for comparison.}
\label{fig3}
\end{figure}

It is convenient to represent the total Helmholtz free energy
$F ( V , T )$ of the system at volume $V$ and temperature $T$ as the
sum of three parts:~\cite{alfe01}
\begin{equation}
F ( V , T ) = F_{\rm p} ( V , T ) + F_{\rm h} ( V , T ) +
F_{\rm a} ( V , T ) \; .
\label{eq:freeener}
\end{equation}
Here, $F_{\rm p}$ is the Helmholtz free energy of the static perfect
lattice: it is a {\em free} energy, because we include thermal
electronic excitations.~\cite{mermin65,alfe01} The second term $F_{\rm h}$
is the free energy due to lattice vibrations, calculated in the
harmonic approximation. The remainder $F_{\rm a}$ accounts for
anharmonicity; we ignore $F_{\rm a}$ here, but show how to compute it
in Sec.~\ref{sec:anharmonic}.

The calculation of $F_{\rm p} ( V , T )$ is completely standard.
It is known from previous work that at $T = 0$~K 
and pressures $P < 660$~GPa the most stable phase of Mo is bcc.~\cite{belonoshko08,cazorla08} 
At higher compressions, Mo stabilizes in the double hexagonal closed-packed
(dhcp) structure as recently shown by Belonoshko \emph{et al.} 
(see Fig.~1 of Ref.~[\onlinecite{belonoshko08}]) 
and confirmed in our own calculations.     
Since we focus here on pressures $P < 600$~GPa, we have 
computed $F_{\rm p} ( V , T )$ on a grid of $( V , T )$ points
for the bcc, fcc and hcp structures.
This grid spans the ranges 
$8.25 \le V \le 15.55$~\AA$^{3}$/atom and $0 \le T \le 10000$~K,
with state points taken at intervals of 
$0.5$~\AA$^{3}$/atom and $500$~K, respectively. 
We then fit the $F_{\rm p} ( V , T )$ results obtained at fixed $T$
to a third-order Birch-Murnaghan equation~\cite{birch78} of the form
\begin{eqnarray} 
F_{\rm p}( V , T ) & = & E_{0} + \frac{3}{2}~V_{0}~K_{0} 
\bigg [ -\frac{\chi}{2} \left ( \frac{V_{0}}{V} \right )^2 + 
\frac{3}{4}~ \left ( 1+2 \chi \right ) 
\left ( \frac{V_{0}}{V} \right )^{4/3} \nonumber \\
& - & \frac{3}{2} \left ( 1+\chi \right ) 
\left (\frac{V_{0}}{V} \right )^{2/3} + \frac{1}{2} 
\left (\chi+\frac{3}{2}\right ) \bigg ] \; ,
\label{eq:eqstate}
\end{eqnarray}
where $E_{0}$ and $K_0 = -V_0 d^2E / dV^2$ are the values of the
energy and the bulk modulus at equilibrium volume $V_{0}$, respectively,
$\chi = \frac{3}{4}\left ( 4 - K_0^\prime \right )$ and 
$K_0^\prime = \left[ \partial K / \partial P \right]$, 
with derivatives evaluated 
at zero pressure. Finally, the dependence of parameters $E_{0}$, $K_{0}$, 
$V_{0}$ and $K_0^\prime$ on $T$ is fitted 
to $4$-th order polynomial expressions.    

We obtain the harmonic phonon frequencies $\omega_{{\bf q}, s}$ by
diagonalising the dynamical matrix, which is the spatial Fourier
transform of the force-constant matrix. Our calculations of the latter
by the small-displacement method,~\cite{kresse95,phon} used large supercells 
of $216$ atoms ($4\times 4\times 4$ $k$-point grid) 
for the bcc and fcc structures and $200$ atoms for hcp 
($4\times 4\times 3$ $k$-point grid).
We performed extensive tests for Mo in the bcc structure which
showed that these parameters~\cite{cazorla07} guarantee $F_{\rm h}$ 
values converged to less than $1$~meV/atom; these parameters are 
assumed to be equally adequate for the fcc and hcp structures. 
In principle, the force-constant matrix and the frequencies 
$\omega_{{\bf q}, s}$ depend on
the electronic temperature, but we ignore this dependence here.
Phonon calculations performed with an electronic $T$ equal to $2000$ and 
$5000$~K provide $F_{\rm h}$ results that agree within 
$1$~meV/atom, so we used $T = 2000$~K in all the 
$\omega_{{\bf q}, s}$ calculations. 

The phonons are stable for bcc over the entire range $0 < P < 600$~GPa,
as is known from previous work~\cite{belonoshko08,zeng10}. 
However, the phonons for fcc and hcp 
are stable only above a threshold pressure $P_{\rm th}$. To
illustrate this, we show in Fig.~\ref{fig1} the fcc phonon frequencies
at $V = 9.64$~\AA$^{3}$/atom ($P = 328$~GPa) and 
$V = 10.19$~\AA$^{3}$/atom ($P = 265$~GPa). We see that
the phonon instability first occurs at a finite wavevector, which we
estimate as ${\bf q}_{\rm inst} = \left( 2\pi/a_{0}\right) 
\left(1/4, 1/4, 0 \right)$. We find threshold
pressures of $P_{\rm th} = 310$ and $325$~GPa for the fcc 
($V = 9.78$~\AA$^{3}$/atom) and hcp ($V = 9.69$~\AA$^{3}$/atom) structures.

When calculating the harmonic vibrational free energy $F_{\rm h} ( V , T )$,
we use the classical expression:
\begin{equation}
F_{\rm h} ( V , T ) = 3 k_{\rm B} T 
\ln ( \hbar \bar{\omega} / k_{\rm B} T ) \; ,
\label{eq:fharm}
\end{equation}
where $\bar{\omega}$ is the geometric mean frequency, defined by:
\begin{equation}
N_{{\bf q}, s}^{-1} \sum_{{\bf q}, s}
\ln ( \omega_{{\bf q}, s} / \bar{\omega} ) = 1 \; ,
\label{eq:geomfreq}
\end{equation}
with the sum going over all $N_{{\bf q}, s}$ phonon modes 
(wavevector ${\bf q}$, branch $s$) in the first Brillouin zone. This
classical formula for $F_{\rm h}$ is valid at temperatures well
above the Debye temperature, which for Mo 
is around $400$~K at equilibrium. We have checked that even at $T = 1000$~K
the difference between the $F_{\rm h}$ values obtained
with the classical and quantum formulas is
less than $1$~meV/atom, so our choice
does not affect the accuracy of the results. 

We show the calculated mean frequencies $\bar{\omega}$ 
for all the structures in Fig.~\ref{fig2}. 
Comparison of the $\bar{\omega}$ values indicates that  
harmonic contributions to the free-energy tend to stabilize 
the fcc and hcp structures over bcc, since   
$\bar{\omega}^{\rm hcp} < \bar{\omega}^{\rm bcc}$ 
and $\bar{\omega}^{\rm fcc} < \bar{\omega}^{\rm bcc}$ in the $V$ range studied. 
The stabilisation is slightly greater for hcp than for fcc.

We fit the dependence of the quantity $\ln ( \hbar \bar{\omega} )$ on $V$ 
to a 3rd-order polynomial for all the structures in order to   
know the value of $F_{\rm h}$ at any $( V , T)$ thermodynamic state
using formula~(\ref{eq:fharm}). 
From the harmonic free energies 
$F^\prime ( V , T ) = F_{\rm p} ( V , T ) + F_{\rm h} ( V , T )$,
we have determined the transition bcc-fcc and bcc-hcp pressures at each 
temperature using the double-tangent construction.
The phase boundaries given by these calculations are shown in 
Fig.~\ref{fig3}, where we also indicate the harmonic phase boundaries 
from the calculations by Belonoshko \emph{et al.}~\cite{belonoshko08} 
(The boundaries given by the very recent quasiharmonic calculations
of Zeng {\em et al.}~\cite{zeng10} are similar.) 
In fact, there are some discrepancies. For instance, those 
calculated by Belonoshko \emph{et al.} lie at somewhat lower $T$ 
than ours, and the slopes of the two harmonic 
bcc-hcp boundaries differ in sign. Despite these differences, 
we agree with Belonoshko \emph{et al.} that in the harmonic approximation
the stable high-$P$/high-$T$ structure of Mo is hcp.

\section{Vibrational, elastic and thermodynamic stability}
\label{sec:stability}

\begin{figure}
\centerline
        {\includegraphics[width=0.8\linewidth]{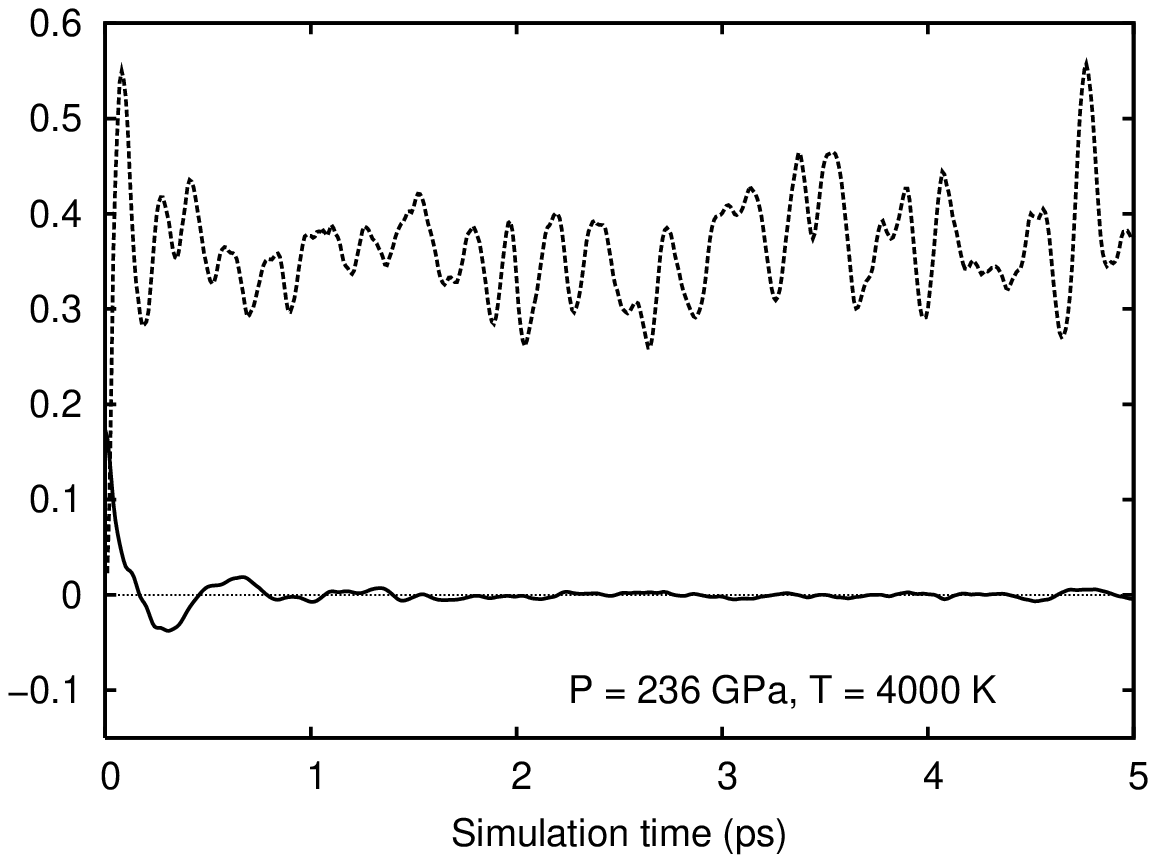}}%
        {\includegraphics[width=0.8\linewidth]{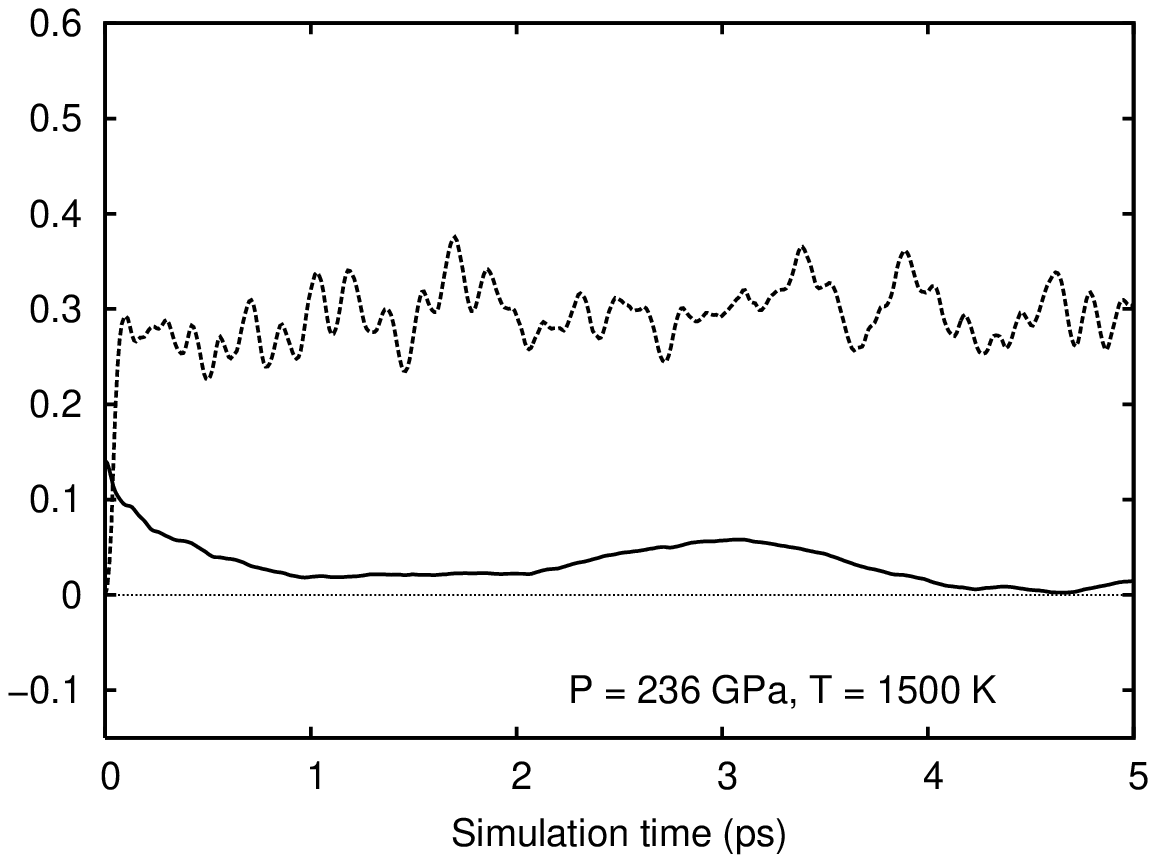}}%
        \caption{Calculated mean squared displacement $\Delta r ( t )^{2}$
         (dashed line) and position correlation function $p ( t )$ (solid line) 
         of Mo in the fcc structure as a function of time. 
         The simulations were performed at two different 
         temperatures and fixed volume $V = 10.50$~\AA$^{3}$/atom. 
         The value of functions $\Delta r ( t )^{2}$ and $p ( t )$ is in units 
         of \AA$^{2}$.}
\label{fig4}
\end{figure}

Do the phase boundaries predicted by harmonic theory have anything
to do with the transitions seen in DAC and shock experiments?
If they do, then the simplest hypothesis is that these transitions
lie on a continuation of the predicted boundaries. But in order for
this to be true, several conditions must be satisfied. First,
since the lower shock transition occurs at $P = 220$~GPa, which is
far below the harmonic stability limit for both fcc and hcp,
the system must somehow be vibrationally stabilised, presumably
by anharmonic effects. Second, the system must remain elastically
stable at pressures $P < 220$~GPa, i.e. small, arbitrary volume-conserving
strains must not cause the free energy to decrease. Third, the crystal
structures must have lower free energies than bcc. Vibrational
stability can be tested by straightforward first-principles MD 
simulations, as has been shown in our earlier work on high-$P$/high-$T$ 
bcc Fe,~\cite{vocadlo03} and in recent work by Asker \emph{et al.} on low-$P$ fcc 
Mo;~\cite{belonoshko08b} we report tests here for fcc Mo. The strain
dependence of free energy can be also probed by MD calculations
in which the thermal average stress is monitored.
For the case of fcc, calculation of stress as a function of strain
along the Bain path also allows us to test its thermodynamic stability.

\subsection{Vibrational stability}
\label{sec:vibrational_stability}

When we say that a crystal in thermal equilibrium is vibrationally
stable, we mean that the thermal average position of each atom
remains centred on its pefect-lattice site, and does not acquire
a permanent deviation away from that site. To test this, it is convenient
to use the so-called position correlation function $p ( t )$,
defined by:~\cite{vocadlo03}
\begin{equation}
p ( t ) = \langle ( {\bf r}_i ( t + t_0 ) - {\bf R}_i^0 ) \cdot
( {\bf r}_i ( t_0 ) - {\bf R}_i^0 ) \rangle \; ,
\end{equation}
where ${\bf r}_i ( t )$ is the position of atom $i$ at time $t$,
${\bf R}_i^0$ is the perfect-lattice position of the atom, $t_0$ is
an arbitrary time origin, and $\langle \, \cdot \, \rangle$ denotes
the thermal average. In practice, the thermal average is performed
by averaging over $t_0$ and over atoms. At $t = 0$, $p ( t )$ is simply
the vibrational mean square displacement. 
The crystal is vibrationally
stable if $p ( t \rightarrow \infty ) = 0$, because the vibrational
displacements at widely separated times become uncorrelated. But if
atoms acquire a permanent vibrational displacement, then 
$p ( t \rightarrow \infty )$ becomes non-zero. The characteristic
behaviour of $p ( t )$ in a vibrationally unstable crystal
can be seen in our earlier work on high-$P$/high-$T$ Fe in the bcc 
structure.~\cite{vocadlo03}
For this test to work, the atoms must not diffuse from one site to another,
and we routinely test for lack of diffusion by monitoring the
time-dependent mean square displacement
$\Delta r ( t )^2 \equiv \langle | {\bf r}_i ( t + t_0 ) - {\bf r}_i ( t_0 )
|^2 \rangle$, which, in the absence of diffusion, goes to a constant
equal to twice the vibrational mean square displacement in the limit
$t \rightarrow \infty$.

We have performed a set of first-principles MD simulations on Mo in
the bcc and fcc structures at a wide range of thermodynamic states.
A typical MD run consisted of $10^{4}$ steps performed with
a time-step of $1$~fs, with the first 5~ps allowed for equilibration,
and only the last 5~ps used to accumulate statistical averages.  
The simulation cell contained $125$ particles for both fcc and bcc structures 
and $\Gamma$-point electronic $k$-point sampling was used.

The MD runs were carried out for a total 
of $20$ state points, spanning the 
ranges $200 \le P \le 600$~GPa and $1500 \le T \le 10500$~K.
In Fig.~\ref{fig4}, we show the mean squared displacement
$\Delta r ( t )^2$ and
the position correlation function $p ( t )$ calculated for fcc Mo 
at volume $V = 10.50$~\AA$^{3}$/atom and $T = 1500$ 
and $4000$~K. At the lower $T$, the system is vibrationally unstable,
as shown by the long-$t$ behaviour of $p ( t )$.
Clearly, atomic liquid-like diffusion does not occur
as shown by the fluctuation of $\Delta r ( t )^{2}$
about a constant value at long $t$. 
Since these MD simulations are very demanding, we did not attempt
to estimate an accurate boundary in the $P - T$ plane separating
stable and unstable states of the fcc structure.
Nevertheless, we can say that vibrational instability
was not observed in our simulations at temperatures 
$T > 3000$~K and pressures below $P_{\rm th} = 310$~GPa.
The recent work of Asker \emph{et al.}~\cite{belonoshko08b}
using techniques similar to those used here, showed that even
at $P \sim 0$~GPa fcc Mo is vibrationally stable for $T \ge 3000$~K.

\subsection{Elastic and thermodynamic stability}
\label{sec:elastic_stability}

\begin{figure}
\centerline{
\includegraphics[width=0.8\linewidth]{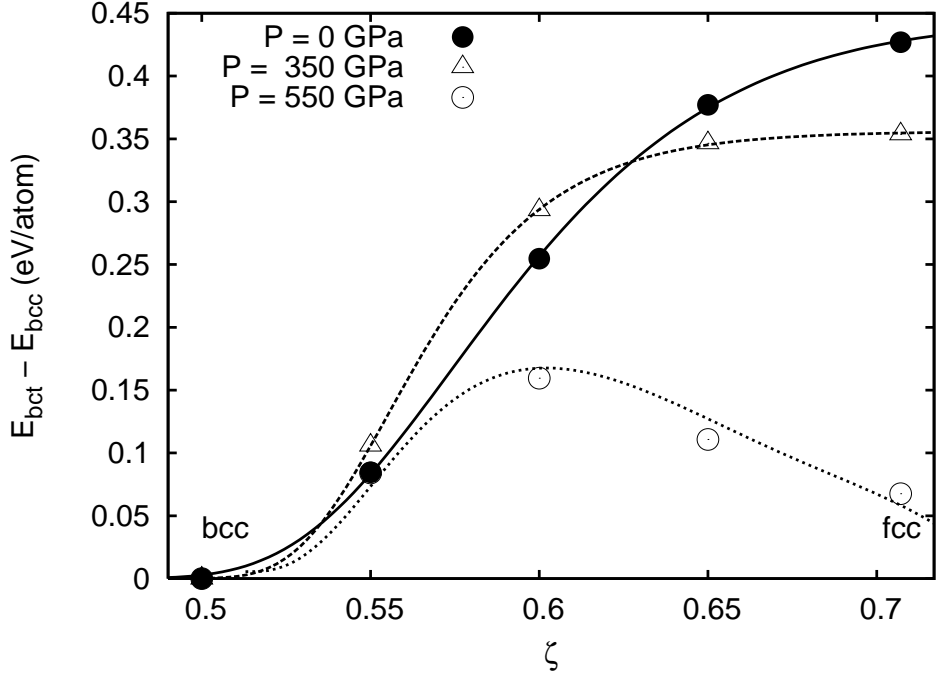}}%
\caption{{\em Ab initio} free energy calculations performed 
         along the Bain path at zero temperature.
         Energy differences are represented with symbols, and 
         lines are guides to the eye.}
\label{fig5}
\end{figure}

\begin{figure}
\centerline
        {\includegraphics[width=0.7\linewidth]{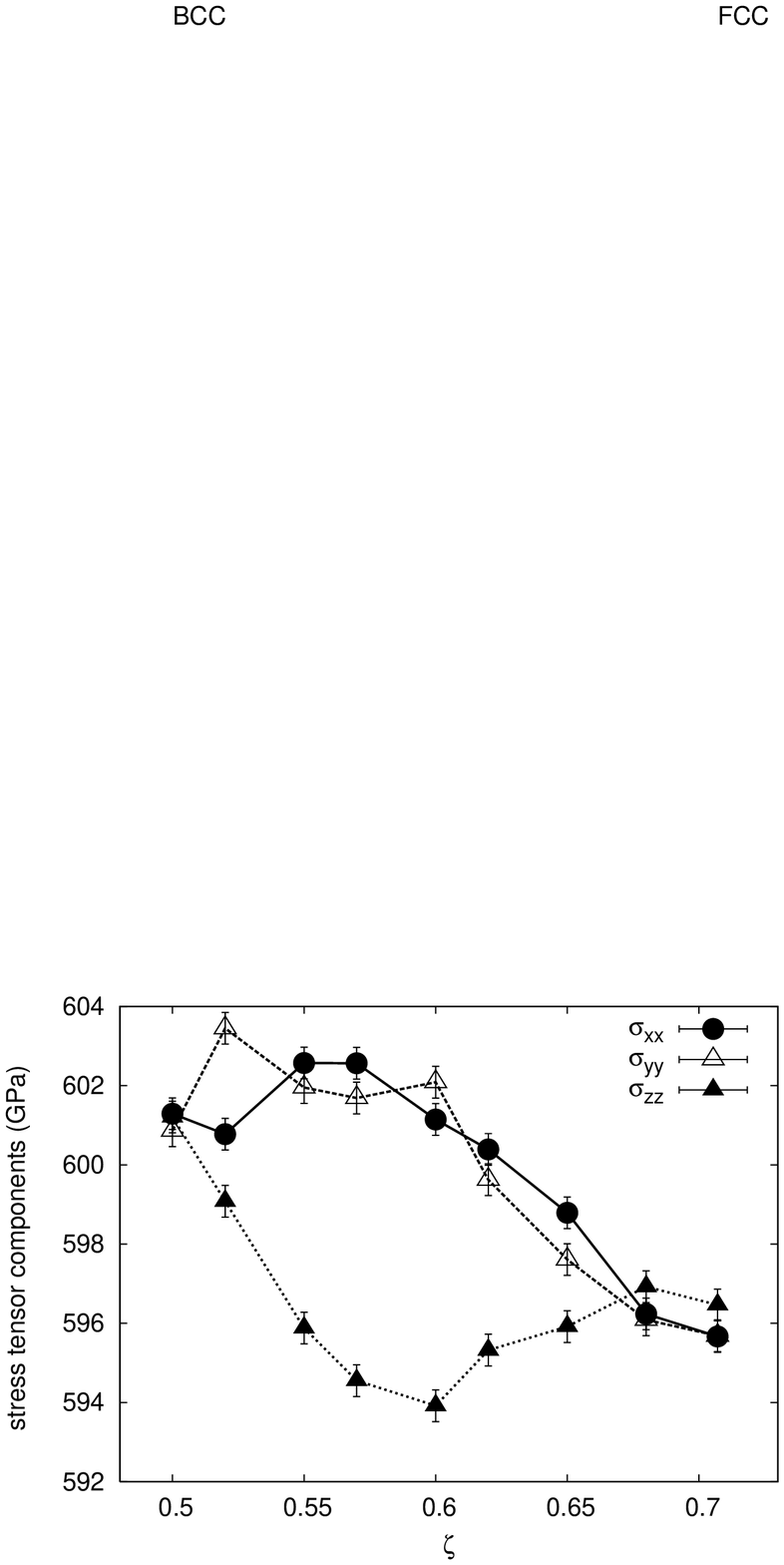}}%
        {\includegraphics[width=0.7\linewidth]{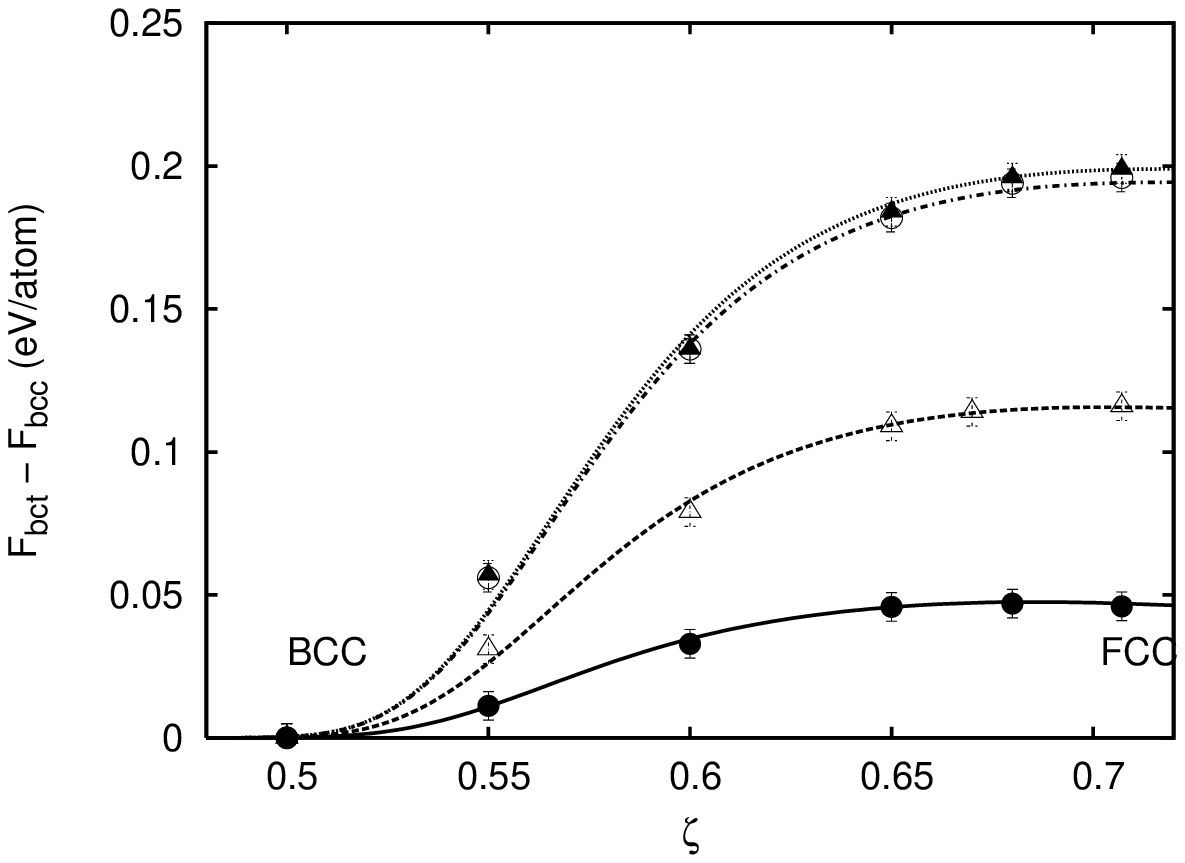}}%
        \caption{{\emph Top}: Stress tensor components calculated at 
         different $\zeta$-points along the Bain path at $V = 8.26$~\AA$^{3}$/atom
         and $T = 9000$~K. Statistical errors are represented with bars 
         equivalent to $0.4$~GPa.
         {\emph Bottom}: Free energy difference $\Delta F ( \zeta )$
         obtained at $( V , T )$ states $( 8.26 , 9000 ) = \bullet$, 
         $( 10.08 , 6000 ) = \triangle$, $( 10.54 , 4000 ) = \circ$ 
         and $( 11.00 , 4000 ) = \blacktriangle$ in units of \AA$^{3}$/atom
         and K, respectively. Numerical uncertainties are represented with bars 
         of $5$~meV/atom and the lines are guides to the eye.}
\label{fig6}
\end{figure}

The elements of the stress tensor $\sigma_{\alpha \beta}$ in a
thermal-equilibrium system can be defined as 
$\sigma_{\alpha \beta} = V^{-1} \left( \partial F \left/ 
\partial \epsilon_{\alpha \beta} \right. \right)_T$, 
where $F$ is the Helmholtz free energy, $\epsilon_{\alpha \beta}$
is the strain tensor, and $V$ is the volume. This relation can
be integrated to obtain the difference of free energy between two
states that differ by a finite homogeneous strain.~\cite{nielsen85,alfe98} 
The bcc and fcc structures can be continuously deformed into one another by such a strain,
following the Bain path. This means that the free energy difference
between the two structures can be obtained by performing a series of
MD simulations along the Bain path, calculating the thermal
average $\sigma_{\alpha \beta}$ in each simulation, and then integrating
numerically with respect to $\epsilon_{\alpha \beta}$. A necessary
condition for elastic stability of the fcc phase is that
$F$ must be a local minimum along the Bain path.~\cite{caveat}

The Bain path is based on the idea that the bcc and 
fcc structures can be regarded as special cases of the 
body-centered tetragonal lattice
(bct, $I4/mmm$ space group). Taking primitive vectors 
${\bf a}_{1} = (1, 0, 0) a$, ${\bf a}_{2} = (0, 1, 0) a$,
${\bf a}_{3} = (1/2, 1/2, \zeta) a$, the values $\zeta = 1/2$ and
and $\zeta = 1/\sqrt{2}$ correspond to bcc and fcc, respectively. 
By varying $\zeta$ from 
$1/2$ to $1/\sqrt{2}$, while varying $a$ so as to  
keep the volume $a^3 \zeta$ of the unit cell constant, 
one structure is transformed continuously into the other. 
If we denote by $F_{\rm bct} ( \zeta )$ the free energy for a given
$\zeta$ value, then the work done on going from the bcc value
$\zeta = 1/2$ to the another value at constant volume is readily shown to be:
\begin{equation}
F_{\rm bct} ( \zeta ) - F_{\rm bcc} =
\frac{1}{3} V \int_{1/2}^\zeta ( \sigma_{xx} + \sigma_{yy} - 2 \sigma_{zz} )
\frac{1}{\zeta^\prime} \, d \zeta^\prime \; .
\label{eqn:integrate_stress}
\end{equation}
For $\zeta = 1 / \sqrt{2}$, we obtain the free energy difference
of interest $F_{\rm fcc} - F_{\rm bcc}$.

As a preliminary test of the correctness of our procedures, we have
performed calculations at $T = 0$~K, in which case the free energy
difference at constant volume is simply the energy difference.
We show in Fig.~\ref{fig5} the 
results of integrating the stress for a range of $\zeta$ values, 
starting from bcc. As expected, at $P = 550$~GPa the difference 
$\Delta E \equiv E_{\rm fcc} - E_{\rm bcc}$
has the small value $0.068$~meV/atom; at $P = 350$~GPa, which is close to the
pressure at which the fcc structure becomes elastically stable, 
$\Delta E$ has the much larger value $0.354$~meV/atom, and
the slope of $\Delta E$ is close to zero at the fcc structure; at
$P = 0$~GPa, the curvature of $\Delta E$ is downwards, so that the fcc
structure is elastically unstable.

Before starting full DFT Bain-path calculations, we have made preparatory
tests to find out how to design the simulations so as to obtain useful
accuracy. These tests were done with an embedded-atom empirical 
potential (EAM),~\cite{daw84,finnis84} which was tuned to reproduce the energetics 
of Mo in the bcc and fcc structures as described by DFT MD simulations 
performed at $V = 9.64$~\AA$^{3}$/atom and $T = 7500$~K. 
The values of the corresponding EAM parameters are, 
with the same notation as in Ref.~[\onlinecite{cazorla07}]~(Eq.~(1)),
$\epsilon = 0.2218$~eV, $a = 5.5525$~\AA, $C = 4.3164$, $n = 3.33$ and $m = 4.68$.    
We set the requirement that integration along the
Bain path should give the free energy difference 
$\Delta F \equiv F_{\rm fcc} - F_{\rm bcc}$ with 
errors of no more than
$\sim 10$~meV due to statistical uncertainty, number of $\zeta$-points
for numerical integration, and system size. Our tests
indicated that the statistical uncertainty in $\sigma_{\alpha \beta}$ should be
less than $\sim 0.5$~GPa, and this is achieved with runs of $3-4$~ps
after equilibration. Using the trapezoidal rule for numerical integration,
we find that nine $\zeta$ points (including the end-points
$\zeta = 1/2$ and $1 / \sqrt{2}$) suffice. 

Guided by the results of these tests, we performed the DFT MD
calculations on systems of $125$ atoms ($\Gamma$-point sampling), 
at nine $\zeta$ values, with an
equilibration time of $2$~ps and a statistical sampling time of $3-4$~ps;
we use our standard plane-wave cut-off of $224.6$~eV. (Checks on
the adequacy of $\Gamma$-point sampling and our standard cut-off
are noted below.)
The Bain-path calculations were done at four different $( V , T )$
states (units of \AA$^3$/atom and K): 
$( 8.26, 9000 )$, $( 10.08, 6000 )$, $( 10.54, 4000 )$
and $( 11.00, 4000 )$, where the pressures corresponding to the bcc structure
are $600$, $283$, $226$ and $187$~GPa, respectively.  
As an illustration, we show in
Fig.~\ref{fig6} the computed values of $\sigma_{xx}$, $\sigma_{yy}$ and 
$\sigma_{zz}$ as a function of $\zeta$ for the state $( 8.26, 9000 )$.
The Figure also reports the free energy difference 
$F_{\rm bct}~(\zeta) - F_{\rm bcc}$ obtained 
by integration at these four states. 

Two important conclusions are clear from these results. First, the
fcc structure is thermodynamically unstable with respect to bcc
at all the high-$T$ states we have examined. This is true even
at the state ($600$~GPa, $9000$~K), which lies on the bcc-fcc boundary
predicted by harmonic theory, and at the state ($283$~GPa, $6000$~K),
which is somewhat above the extension of that boundary. The second
conclusion is that fcc appears to be elastically unstable for the three
states having $P < 300$~GPa, though it is weakly stable at
($600$~GPa, $9000$~K). The conclusions appear to be robust, since they
would be unchanged even if the statistical errors were considerably
greater than those we have achieved. We have tested
for the possible effect of systematic errors coming from 
our use of $\Gamma$-point sampling and our standard plane-wave
cut-off. To test the effect of cut-off, we have repeated the
calculation of the thermally averaged stress components
$\sigma_{xx}$, $\sigma_{yy}$ and $\sigma_{zz}$ at $\zeta = 0.60$ for the
thermodynamic state $V = 8.26$~\AA$^3$/atom and $T = 9000$~K,
using the increased plane-wave cut-off of $280.7$~eV. We find
that this increased cut-off leaves all the stress components
unchanged within $\sim 1$~GPa. We have done the same thing
with the standard cut-off but now using Monkhorst-Pack
$( 2 \times 2 \times 2 )$ sampling of $8$~$k$-points. This has the
effect of shifting all three stress components down by
$\sim 4$~GPa. Since they are all shifted by essentially the same
amount, this does not affect the integral of 
eqn~(\ref{eqn:integrate_stress}), so that
the free-energy difference between fcc and bcc remains 
unchanged.~\cite{paw_radius}

Since the Bain-path calculations fully include anharmonicity, and since
they are completely at odds with the harmonic predictions, it appears
that anharmonic contributions to the free energy must be very
substantial at high temperatures. We examine these contributions
directly in the next Section.

\section{Anharmonic free energy}
\label{sec:anharmonic}

\begin{figure}
\centerline{
\includegraphics[width=0.8\linewidth]{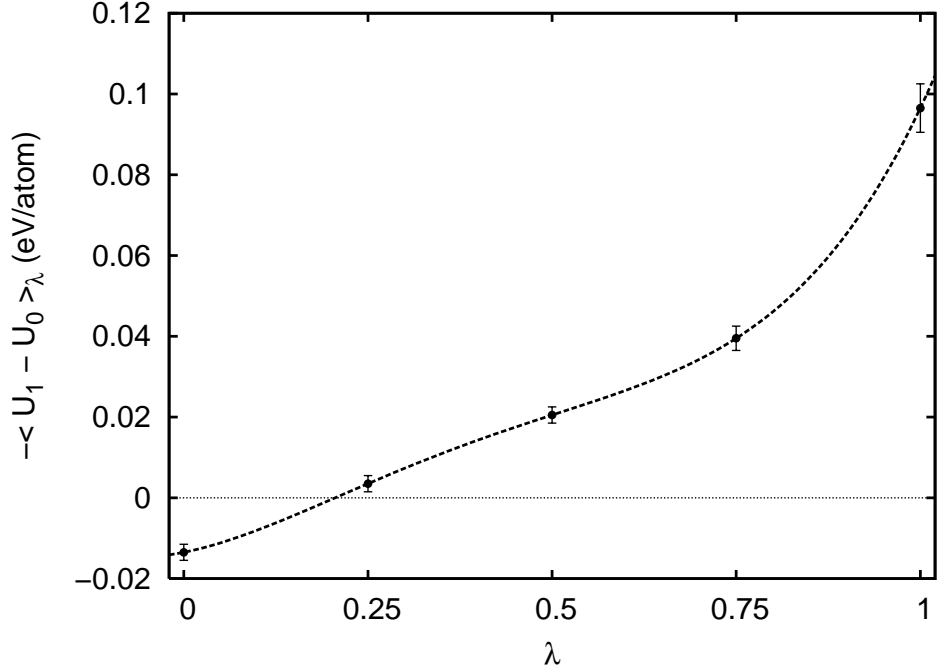}}%
\caption{Averaged $\langle U_{1} - U_{0}\rangle_{\lambda}$ values obtained 
         at different $\lambda$-points in anharmonic free energy
         calculations perfomed for bcc Mo at $V = 10.08$~\AA$^{3}$/atom
         and $T = 6000$~K. The dashed line corresponds to a 4-th order polynomial
         curve used to reproduce the variation of these values on parameter $\lambda$.}
\label{fig7}
\end{figure}

\begin{figure}
\centering
{\includegraphics[width=0.70\linewidth]{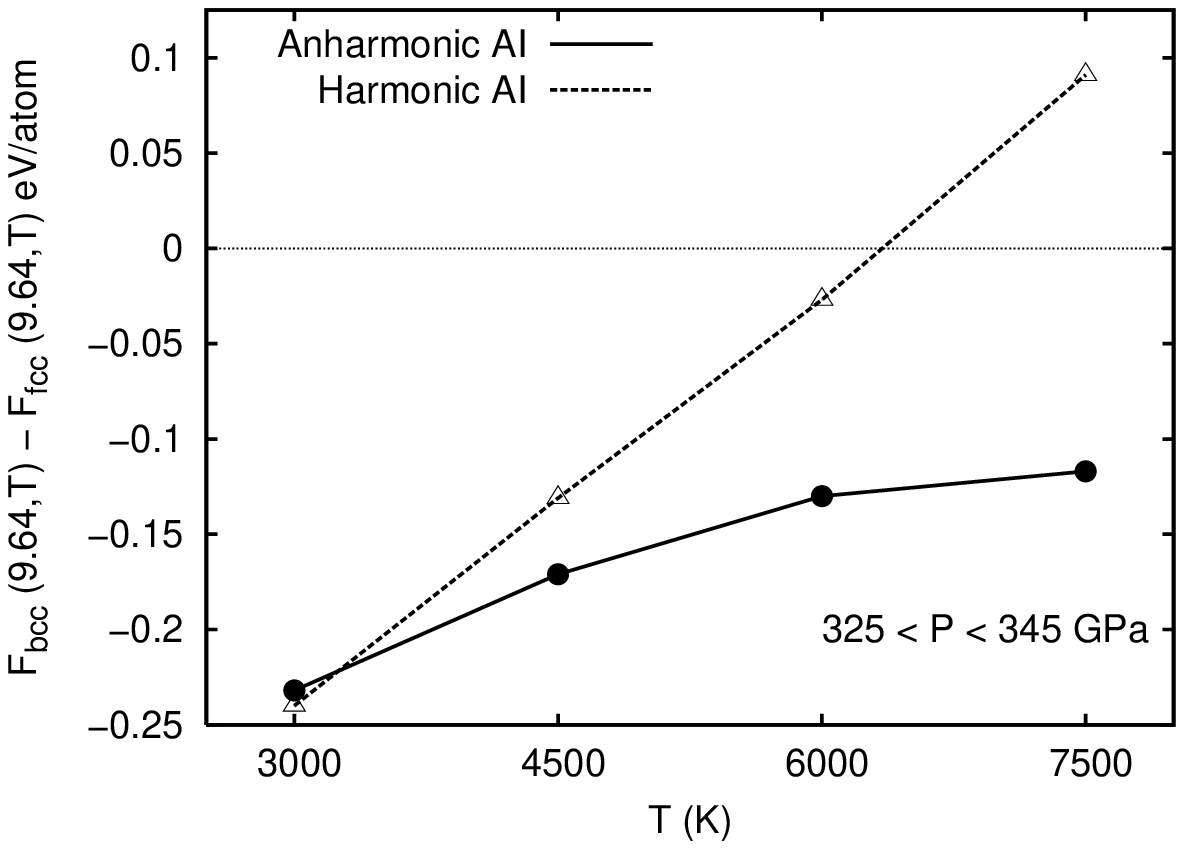} }%
{\includegraphics[width=0.70\linewidth]{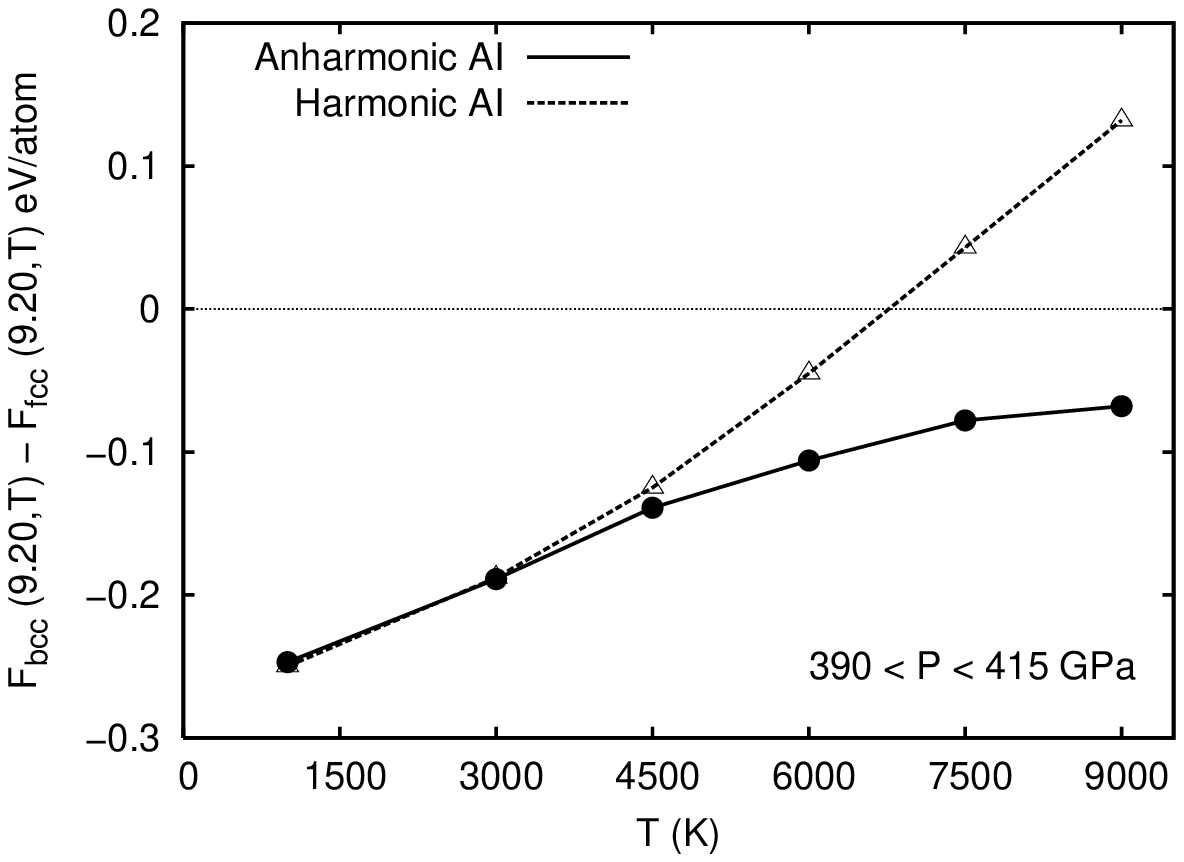} }%
\vspace{0.0cm}
\caption{Dependence on temperature of the free energy difference between bcc and fcc Mo
         as given by anharmonic and quasi-harmonic DFT free energy calculations performed at
         two different volumes.}
\label{fig8}
\end{figure}

We calculate the anharmonic contribution to the free energy using
thermodynamic integration, which we have used extensively in previous
work on the free energy of transition metals.~\cite{alfe02a,alfe02b,alfe01} 
The general principle is that we compute the change of Helmholtz free energy as the total
energy function $U_\lambda ( {\bf r}_1 , \ldots {\bf r}_N )$ is
changed continuously from $U_0$ to $U_1$, the free energies associated
with these energy functions being $F_0$ and $F_1$. Then the
thermodynamic integration formula is:
\begin{equation}
F_1 - F_0 = \int_0^1 \langle U_1 - U_0 \rangle_\lambda \, d \lambda \; ,
\end{equation}
where $\langle \, \cdot \, \rangle_\lambda$ is the thermal average evaluated
for the system governed by the energy function 
$U_\lambda = ( 1 - \lambda ) U_0 + \lambda U_1$. In practice, we take
$U_1$ to be the DFT total energy function $U$, whose free energy we wish
to calculate, and $U_0$ to be the total energy function $U_{\rm ref}$
of a ``reference'' system, chosen so that its free energy $F_{\rm ref}$
can be evaluated exactly. Here, we choose the reference system
to be a perfectly harmonic system~\cite{harm_ref}. For 
a volume where the harmonic
phonons are all stable, we can choose $U_{\rm ref}$ to be the total energy
of the DFT system calculated in the harmonic approximation.
For $V$ where DFT gives imaginary phonon frequencies, the total free
energy cannot be separated into perfect-lattice, harmonic and
anharmonic components (Eq.~(\ref{eq:freeener})). However, we know 
from Sec.~\ref{sec:vibrational_stability}
that the fcc (and hcp) system can still be vibrationally
stable at such volumes, so that the free energy should still
be calculable. In these cases, we create a harmonic reference system
by adding artificial on-site harmonic springs to remove the
harmonic instability.

In applying this scheme in practical DFT calculations, there is a subtle
point connected with electronic $k$-point sampling, which we note here.
Ideally, we should use infinitely fine $k$-point sampling; we denote
the DFT total energy calculated in this way by
$U^\infty ( {\bf r}_1 , \ldots {\bf r}_N )$. (As usual, $U^\infty$ is
a {\em free} energy, because it includes thermal electronic excitations.)
However practical DFT simulations have to be performed with limited
$k$-point sampling, and we denote the total energy in this case
by $U^k ( {\bf r}_1 , \ldots {\bf r}_N )$. (In fact, most of
our simulations are performed with $\Gamma$-point sampling.) Now with
both perfect and imperfect $k$-point sampling, we can separate
the total energy into the total (free) energy of the perfect lattice
$U_{\rm p}$ and the vibrational energy $U_{\rm vib}$. We write
$U^\infty = U_{\rm p}^\infty + U_{\rm vib}^\infty$ and
$U^k = U_{\rm p}^k + U_{\rm vib}^k$. Now the energies
$U_{\rm p}^\infty$ and $U_{\rm p}^k$ are very large, and we do not
wish to incur $k$-point errors in these; we do not need to do so,
since $U_{\rm p}^\infty$ and $U_{\rm p}^k$ can be calculated
explicitly in advance. In a practical DFT simulation with
limited $k$-point sampling, we therefore make smaller errors
if we take the total energy to be $U_{\rm p}^\infty + U_{\rm vib}^k =
U_{\rm p}^\infty + ( U^k - U_{\rm p}^k )$. The reference system
should be taken to have the total energy
$U^{\rm ref} = U_{\rm p}^\infty + U_{\rm h}^{\rm ref}$, where
$U_{\rm h}^{\rm ref}$ is a bilinear function of the displacements
of atoms from their regular lattice sites.

With these points in mind, the $\lambda$-dependent total energy
function used in thermodynamic integration is:
\begin{equation}
U_\lambda = U_{\rm p}^\infty + ( 1 - \lambda ) U_{\rm h}^{\rm ref} +
\lambda ( U^k - U_{\rm p}^k ) \; .
\end{equation}
The total free energy of the system is then:
\begin{equation}
F = U_{\rm p}^\infty + F_{\rm h}^{\rm ref} +
\int_0^1 d \lambda \,
\langle U^k - U_{\rm p}^k - U_{\rm h}^{\rm ref} \rangle_\lambda \; .
\label{eq:freener}
\end{equation}

We evaluate the integral in Eq.~(\ref{eq:freener}) numerically. For this, 
first we perform a series of \emph{ab initio} molecular dynamics simulations
in the $( N , V , T )$ ensemble governed by the energy function $U_{\lambda}$ 
at different $\lambda$-values. 
We then fit a fourth-order polynomial to the 
$\langle U^k - U_{\rm p}^k - U_{\rm h}^{\rm ref} \rangle_{\lambda}$ 
values obtained from these simulations, in order to perform  
the $\lambda$-integration. Our tests show that  
DFT MD simulations performed at five equally spaced 
$\lambda$-points are enough   
to ensure convergence of the free energy to better than $10$~meV/atom
(see Figure~\ref{fig7}). 
A typical run consisted of $3\times 10^{3}$ MD steps performed with
an time-step of $1$~fs with statistical averages taken over the 
last $2$~ps. This procedure gives values for
$\langle U^k - U_{\rm p}^k - U_{\rm h}^{\rm ref} \rangle_{\lambda}$ 
converged to better than $6$~meV/atom.  
The simulation box employed contains $125$ particles 
($128$ in the hcp case)
and $\Gamma$-point electronic sampling was used.

\begin{table}
\begin{center}
\label{tab:}
\begin{tabular}{c c c c c c c c}
\hline
\hline
$ ~V~ $&$ ~T~ $&$ F^{bcc} $&$ F^{bcc}_{a}~$&$ F^{fcc}$&$  F^{fcc}_{a}~$&$ F^{hcp} $&$ F^{hcp}_{a} ~$\\ 
\hline
 $9.20$&$1000$&$-5.694$ &$-0.002 $&$-5.447$ &$-0.005$&$ $ &$ $\\
 $( 390 \le P \le 415 )  $&$3000$&$-7.014$ &$-0.005$&$-6.825$ &$-0.004$&$-6.619$&$-0.015$\\
 $    $&$4500$&$-8.295$ &$0.006$&$-8.156$ &$0.020$&$-7.964$&$0.003$\\
 $    $&$6000$&$-9.747$ &$0.001$&$-9.641$ &$0.062$&$-9.472$&$0.022$\\
 $    $&$7500$&$-11.336$&$-0.017$&$-11.258$&$0.104$&$$&${\rm unstable}$\\
 $    $&$9000$&$-13.047$&$-0.051$&$-12.979$&$0.149$&$ $ &$ $\\
\hline
 $9.64$&$3000$&$-8.117$ &$0.000$&$-7.885$ &$-0.008$&$ $ & $  $\\
 $( 325 \le P \le 345 )  $&$4500$&$-9.409$ &$-0.005$&$-9.238$ &$0.035$&$ $ & $  $\\
 $    $&$6000$&$-10.872$&$-0.005$&$-10.742$&$0.098$&$ $ & $  $ \\
 $    $&$7500$&$-12.484$&$-0.038$&$-12.367$&$0.170$&$ $ & $  $ \\
\hline
$10.08$&$3000$&$-9.031 $ & $0.005$ & $-8.783 $ & $ $ & $ $ & $ $  \\
$( 270 \le P \le 285 )   $&$4500$&$-10.338$ & $0.001$ & $-10.160$ & $ $ & $ $ & $ $ \\
$     $&$6000$&$-11.819$ & $-0.019$ & $-11.691$ & $ $ & $ $ & $ $ \\
\hline 
\hline
\end{tabular}
\end{center}
\caption{Total and anharmonic free energy values obtained for Mo in 
         different crystal structures within the thermodynamic range       
         $270 \le P \le 415$~GPa and $ 3000 \le T \le  9000$~K. 
         The cases in which the value of the $F_{a} = F - F_{p} - F_{h}$ 
         term is not shown correspond to thermodynamic states 
         at which the corresponding crystal structure is 
         harmonically unstable. Volumes are in units of \AA$^{3}$/atom,   
         pressures of GPa and free energies of eV/atom.} 
\end{table}

Anharmonic free energy results are shown in Table~I. 
We see that solid Mo is always thermodynamically more stable in 
the bcc structure than in the other structures examined. 
This conclusion disagrees with DFT calculations performed 
in the quasi-harmonic approximation (Sec.~\ref{sec:harmonic}) 
which predict fcc and hcp Mo as more stable than 
bcc Mo at pressures and temperatures above $\sim 350$~GPa and
$\sim 5000$~K. Moreover, the total free energy of Mo in the hcp 
phase is around $\sim 0.1$~eV/atom larger than in the bcc or 
fcc structures. The reason behind these results is that the anharmonic 
energy term $F_{a}$ in general is negative for the bcc structure while 
positive for the rest of structures, particularly at low pressures and high 
temperatures (see Table~I and Fig.~\ref{fig8}).
We find very good agreement between  
the results obtained using thermodynamic integration (TI) and integration 
of the stress tensor with respect to strain along the Bain path (BP);
for instance, at state $( 10.08 , 6000 )$ the free energy difference 
$F_{bcc} - F_{fcc}$ obtained with TI is $-0.128$~eV/atom 
while the corresponding BP value is $-0.117$~eV/atom.

The results of the present Section and the previous one all indicate
that neither fcc nor hcp becomes thermodynamically more stable than bcc
at high temperature. If this is true, then the melting curves of
those two crystal structures should lie lower in temperature than the
bcc curve. We turn to this question for fcc Mo in the next Section.

\section{Melting curve of fcc Mo}
\label{sec:melting}

\begin{figure}
\centerline{
\includegraphics[width=0.8\linewidth]{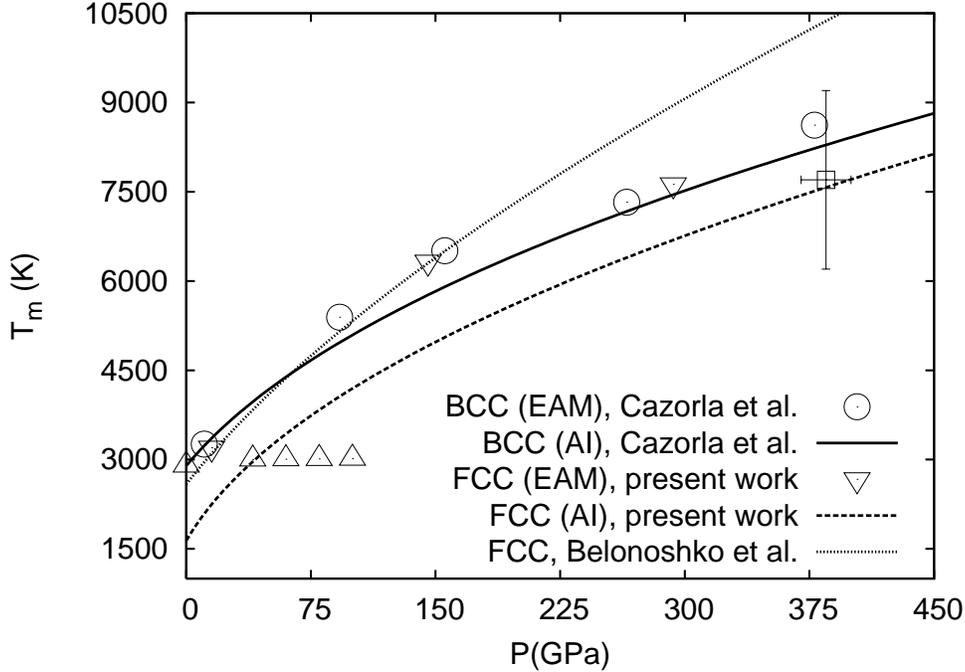}}%
\caption{\emph{Ab initio} (AI) high-$P$ and high-$T$ melting curve of Mo calculated
         for the bcc~[\onlinecite{cazorla07}] (Cazorla \emph{et al.}) and fcc
         (present work) crystal structures. 
         Melting ($P_{m}^{\rm ref}$, $T_{m}^{\rm ref}$) 
         states obtained in the two-phase coexistence simulations performed with 
         EAM potentials are also displayed. $\triangle$~[\onlinecite{errandonea01}]
         and $\Box$~[\onlinecite{hixson92}] represent DAC and shock-wave data, respectively. The melting line of
         fcc Mo as calculated by Belonoshko \emph{et al.}~[\onlinecite{belonoshko08}] is shown for comparison.}
\label{fig9}
\end{figure}

There are several well established techniques for calculating first-principles
melting curves,~\cite{gillan06} including the 
calculation of the free energies of solid
and liquid using thermodynamic integration from reference systems,
and the direct first-principles simulation of coexisting solid
and liquid in large systems. Here, we 
begin (Sec.~\ref{subsec:ref_coexist}) by using the ``reference coexistence''
method,~\cite{cazorla07,taioli07,vocadlo04,alfe04,alfe02a}, 
because it is fairly easy to apply 
and because we have used it recently to determine the DFT 
melting curve of bcc Mo. We shall see that the results given by this method
are inconsistent with earlier results for
the relation between the melting curves of bcc
and fcc Mo obtained by the Z method.~\cite{belonoshko08} In order to
investigate the reasons for this discrepancy, we shall present
(Sec.~\ref{subsec:Z_method}) our own DFT Z-method calculations,
using larger simulated systems and longer simulation times than
were used in the earlier work.

\subsection{Reference coexistence}
\label{subsec:ref_coexist}

The reference coexistence technique consists of three steps. First,
an empirical reference model is fitted to first-principles
simulations of solid and liquid at thermodynamic conditions close
to the expected melting curve. Next, the reference model is used
to perform simulations of coexisting solid and liquid in large systems
consisting of many thousands of atoms, so as to find points
$( P_{\rm m}^{\rm ref} , T_{\rm m}^{\rm ref} )$ on the reference
melting curve. Finally, differences between first-principles
and reference free energies of the solid and liquid are used to estimate
the differences between reference and first-principles melting curves.
In the case of Fe, reference coexistence results have been compared
with melting curves obtained both by first-principles free energy
calculations and by direct first-principles simulation of coexisting
solid and liquid, and the agreement was excellent.~\cite{alfe02b,alfe09}
Moreover, notable agreement between reference coexistence
results and diffusion Monte Carlo free energy calculations has been 
also proved recently.~\cite{sola09} 

The reference model used in our reference coexistence calculations
on the melting of bcc Mo was an embedded atom model (EAM), details
of which are given in Ref.~[\onlinecite{cazorla07}]. 
We use exactly the same model
with the same parameters here. In our work on bcc Mo, we showed that
EAM coexistence simulations on cells containing 6750 atoms give
accurate reference melting curves, and we use the same size of system here.
The protocols used to prepare the two-phase system are the same as
those used before, and we accept a thermodynamic state
$( P_{\rm m}^{\rm ref} , T_{\rm m}^{\rm ref} )$ as lying on the
reference melting curve if the two phases remain in stable
coexistence for 50~ps or more. The reference melting curve obtained
for fcc Mo in the present work is compared with our published reference curve
for bcc Mo in Fig.~\ref{fig9}. The two curves are essentially identical.

The leading-order shift $\Delta T_{\rm m}$ in melting temperature
caused by going from the reference to the first-principles
total-energy function is:
\begin{equation}
\Delta T_{\rm m} = \Delta G^{l s} ( T_{\rm m}^{\rm ref} ) /
S_{\rm ref}^{\rm l s} \; .
\end{equation}
Here, $\Delta G^{l s} \equiv \Delta G^l -\Delta G^s$, where
$\Delta G^l$ and $\Delta G^s$ are the isobaric-isothermal changes
of Gibbs free energy of liquid and solid due to the change
$\Delta U$ of total-energy function; the denominator
$S_{\rm ref}^{l s}$ is the reference entropy of fusion, i.e.
the difference between the entropies of liquid and solid in the
reference model. The free energy shifts $\Delta G^l$ and
$\Delta G^s$ are calculated using the formula:
\begin{equation}
\Delta G = \langle \Delta U \rangle_{\rm ref} -
\frac{1}{2} \beta \langle \delta \Delta U^2 \rangle_{\rm ref} -
\frac{1}{2} V \kappa_T \Delta P^2 \; ,
\label{eq:gibbshift}
\end{equation}
with $\beta = 1 / k_{\rm B} T$, 
$\delta \Delta U \equiv \Delta U - \langle \Delta U \rangle_{\rm ref}$
(averages taken in the reference system), $\kappa_T$ is the isothermal
compressibility and $\Delta P$ is the isochoric-isothermal
difference of pressure between first-principles and reference systems.

Following the procedures used in our work on bcc Mo, we evaluated
$S_{\rm ref}^{l s}$ and the reference $\kappa_T$ values for solid
and liquid using separate solid- and liquid-state simulations on cells
of 3375 atoms at $( P , T )$ points on the reference melting curve. The
values of $\langle \Delta U \rangle_{\rm ref}$, 
$\langle \delta \Delta U^2 \rangle_{\rm ref}$ and $\Delta P$
were obtained from solid- and liquid-state simulations on 
systems of 125 atoms, using a $2 \times 2 \times 2$ Monkhorst-Pack
grid for electronic $k$-point sampling.

In Fig.~\ref{fig9}, we compare the resulting DFT melting curve for fcc Mo with
the DFT curve for bcc Mo obtained using exactly the same procedures;
we also show the fcc melting curve of 
Belonoshko {\em et al.}~\cite{belonoshko08}
obtained using the Z method.~\cite{belonoshko06} 
We see that the free energy corrections
cause a downward shift of the melting curve for both bcc and fcc,
but the shift is considerably greater for fcc. 
Consequently, the fcc melting curve lies below the bcc curve. 
In Table~II, we show the values of terms $\langle \Delta U\rangle^{ls}_{\rm ref}$ 
and $\langle\left(\delta \Delta U\right)^{2}\rangle_{\rm ref}$ 
(see Eq.~(\ref{eq:gibbshift})) which are required for the calculation of the 
free energy differences in question.  
The finding that the fcc melting curve lies below the bcc
melting curve means that the free energy of fcc must be higher 
than that of bcc in the high-$T$ region just below the melting curves. 
This confirms the conclusions from our Bain-path and anharmonic calculations.
However, our results are not consistent with those of
Belonoshko {\em et al.}~\cite{belonoshko08}, whose Z-method
calculations indicate that the fcc melting curve lies above
the bcc melting curve.

\begin{table}[c]
\begin{center}
\label{tab:gauss}
\begin{tabular}{@{\hspace{1.0cm}} c@{\hspace{1.0cm}}  c  @{\hspace{0.15cm}}c@{\hspace{1.0cm}}  c  @{\hspace{0.15cm}}c @{\hspace{
1.0cm}}c@{\hspace{1.0cm}} }
\hline
\hline
$ T_{\rm m}^{\rm ref} $ (K)  &  $ \langle \Delta U\rangle^{ls}_{\rm ref} / N~({\rm eV/atom}) $ &
$  $ & $ \frac{1}{2}\beta\langle\left(\delta \Delta U\right)^{2}\rangle_{\rm ref}/N~({\rm eV/atom}) $&
$  $ & $ T_{\rm m}^{\rm AI} $ (K)  \\ \cline{4-4}
$   $  &   $      $  & $  $ &  $\rm{Solid} {\hspace{1.0cm}} \rm{Liquid}$ & $  $ &  $   $  \\
\hline
$3200    $  &   $-0.057(2) $   & $  $  &  $0.024(2) {\hspace{0.45cm}} 0.032(2)  $  &   $     $  &  $2249  $ \\
$6325    $  &   $-0.108(2) $   & $  $  &  $0.044(2) {\hspace{0.45cm}} 0.041(2)  $  &   $     $  &  $4910  $ \\
$7625    $  &   $-0.070(2) $   & $  $  &  $0.015(2) {\hspace{0.45cm}} 0.027(2)  $  &   $     $  &  $6690  $ \\
\hline
\hline
\end{tabular}
\end{center}
\caption{Difference $\langle \Delta U \rangle_{\rm ref}^{l s} \equiv \langle \Delta U \rangle_{\rm ref}^l -
          \langle \Delta U \rangle_{\rm ref}^s$ between the liquid and fcc solid thermal averages
          of the difference
          $\Delta U \equiv U_{\rm AI} - U_{\rm ref}$ of \emph{ab initio} and reference energies,
          and thermal averages in solid and liquid $\langle \left( \delta \Delta U \right)^2\rangle_{\rm ref}$
          of the squared fluctuations of
          $\delta \Delta U \equiv \Delta U - \langle \Delta U \rangle_{\rm ref}$,
          with averages evaluated in the reference
          system and normalized by dividing by the number of atoms $N$.
          Melting temperatures for the reference and \emph{ab initio} systems
          are also reported.}
\end{table}

\subsection{The Z method}
\label{subsec:Z_method}

The electronic-structure methods used in our reference-coexistence
calculations and in the Z-method calculations of 
Belonoshko {\em et al.}~\cite{belonoshko08}
are essentially the same (PAW with the VASP code), so the
contradictory conclusions about the relation between the bcc and fcc
melting curves must originate in differences between the
statistical-mechanical methods. The Z method has been validated by
testing it against known results for the Lennard-Jones and other
systems, using MD simulations on large systems of up to $32,000$  
atoms with simulation times of $\sim 60$~ps~\cite{belonoshko06}. 
However, the DFT 
Z-method simulations of Belonoshko {\em et al.}~\cite{belonoshko08} 
on the melting of Mo employed much smaller systems (from 32 to 108 atoms
for fcc, and from 54 to 128 atoms for bcc), and very short
simulation times of $\sim 3$~ps. It is therefore a natural
question whether the use of such short simulations on such
small systems might be the cause of the discrepancy. We have
very recently investigated in detail the dependence of Z-method
errors on system size and simulation time, and our findings
shed light on this question~\cite{alfe11}. Guided by this, we have performed
our own DFT Z-method calculations on the melting of Mo, and
we report the results here.

The Z method is based on the phenomenon of homogeneous
melting of a superheated solid~\cite{belonoshko06}. The idea is that if
an MD simulation is performed at constant total energy $E$
and volume $V$ (microcanonical ensemble) starting from the
perfect crystal (all atoms on regular-lattice sites), then
after the solid has thermally equilibrated at some temperature
$T_{\rm sol}$ it will subsequently melt only if $T_{\rm sol}$
exceeds a superheating limit $T_{\rm LS}$. Evidence was
presented in Ref.~\onlinecite{belonoshko06} that, 
as the temperature $T_{\rm sol}$
tends to $T_{\rm LS}$ from above, the temperature $T_{\rm liq}$
and pressure $P_{\rm liq}$ of the liquid formed by homogeneous 
melting tend to a point on the melting curve. Our recent investigation
of homogeneous melting~\cite{alfe11} 
focused on the waiting time $\tau_{\rm w}$,
i.e. the time that elapses before the initial solid at temperature
$T_{\rm sol} > T_{\rm LS}$ melts. In order to gather statistics
about $\tau_{\rm w}$, for each system size (number of atoms $N$) with
specified density $N / V$, and for each value of total energy
(equivalently, for each equilibrated solid temperature $T_{\rm sol}$),
we performed several hundred statistically independent simulations
differing only in the random velocities assigned at the start of
the simulation. The key conclusions were that (a)~$\tau_{\rm w}$
is a stochastic quantity having a roughly exponential probability
distribution; (b)~its mean value $\langle \tau_{\rm w} \rangle$
lengthens rapidly as $T_{\rm sol} \rightarrow T_{\rm LS}$, being
roughly proportional to $1 / ( T_{\rm sol} - T_{\rm LS} )^2$;
(c)~$\langle \tau_{\rm w} \rangle$ also increases as the size
of the system decreases, the dependence being roughly $1 / N$.
We noted that if the total simulation time $t_{\rm sim}$ is
much shorter than $\langle \tau_{\rm w} \rangle$, then
melting is unlikely to be observed even when $T_{\rm sol} > T_{\rm LS}$.
This means that if $T_{\rm m}$ is estimated by performing simulations
of fixed length $t_{\rm sim}$ and seeking the lowest $T_{\rm sol}$ and
$T_{\rm liq}$ for which melting is observed, then $T_{\rm liq}$ will
inevitably overestimate $T_{\rm m}$, and the overestimation will become
worse as the system size is reduced. As an indication of the
difficulties, the results of 
Ref.~\onlinecite{alfe11} suggest that, for a transition
metal with a system size of $N = 100$ and a 
simulation time $t_{\rm sim} = 3$~ps
(values similar to those used by Belonoshko 
{\em et al.}~\cite{belonoshko08}), the
overestimation could well be $\sim 2000$~K. Since the overestimation may differ
for different crystal structures, it is clear that the Z method cannot be
used to compare the melting temperatures of different crystal phases
unless large enough systems are simulated for long enough times.

\begin{figure}
\centerline{
\includegraphics[width=0.8\linewidth]{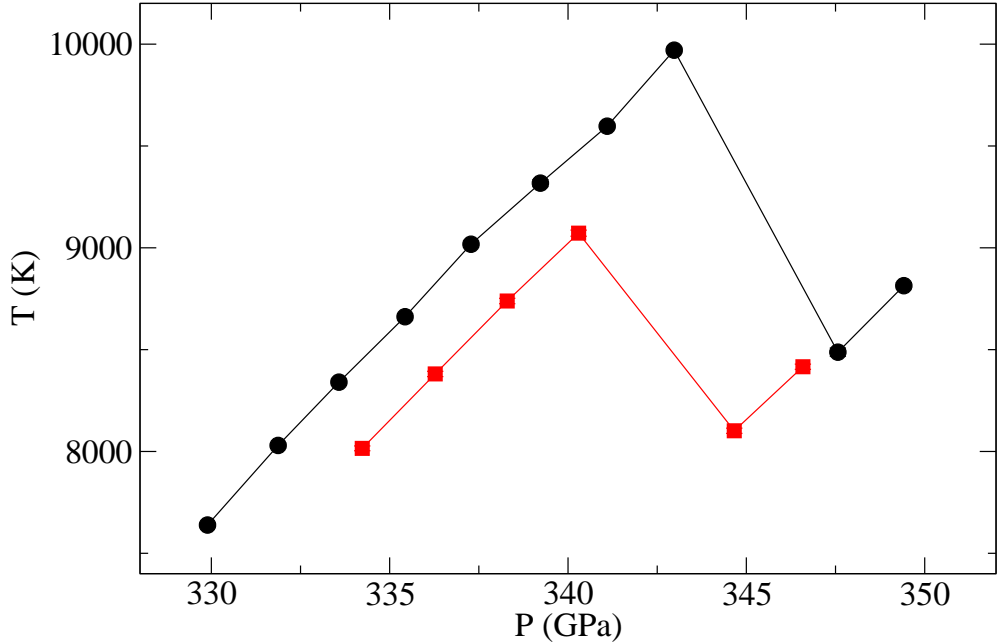}
}
\caption{ Estimation of melting temperatures of bcc and fcc Mo in the
  pressure region P $\simeq 345$ GPa. Round black points and red
  square points show final average (P,T) values from constant energy
  MD simulations (250 atoms for bcc, 256 atoms for fcc) starting from
  the perfect lattice. On the left hand branches, melting does not
  occur within the duration of the simulations (at least 12 ps, see
  text); on the right hand branches melting has occurred, and the
  (P,T) values refer to the liquid.  }
\label{fig_our_Z}
\end{figure}

To illustrate this point, we have performed our own DFT Z-method
simulations on bcc and fcc Mo, using systems of 250 atoms for bcc and
256 atoms for fcc and simulation times of at least $12$~ps (these are
greater than the typical values used in Ref.~\onlinecite{belonoshko08}
by factors of $2.5$ and 4 respectively). The values of the final $T$
and $P$ in our simulations are reported in Fig.~\ref{fig_our_Z}. The
results indicate that the fcc crystal melts at a lower $T$ than bcc,
so that fcc is thermodynamically less stable than bcc, as expected
from our reference-coexistence calculations and from the free energies
from our Bain-path and anharmonic calculations. We note that the
conclusions from the present Z-method simulations are the opposite of
the Z-method results of Ref.~\onlinecite{belonoshko08}. This supports
the suggestion that the earlier Z-method work employed simulations
that were too short on systems that were too small.

\section{Discussion and conclusions}
\label{sec:discussion}

Our results suggest that the fcc and hcp structures
cannot be stable high-$T$ phases of Mo in the pressure range
$0 < P < 600$~GPa. We have shown that they would be more stable
than bcc in the range $350 < P < 600$~GPa and $T > 5000$~K in the
harmonic approximation, as already found by 
Belonoshko {\em et al.}~\cite{belonoshko08} and Zeng {\em et al.}~\cite{zeng10}
However, we find that anharmonic contributions to the free energy
substantially change the picture. Our most direct evidence for this
in the case of fcc comes from thermodynamic integration along the
Bain path, which indicates that fcc is thermodynamically less stable
than bcc in the region where harmonic theory predicts the opposite;
furthermore, fcc appears to be elastically unstable at high~$T$
and $P < 300$~GPa.
The Bain-path approach has the attractive feature that it relies
only on completely standard first-principles MD, and the calculations
are easily repeatable by other researchers. The existence of large
anharmonic contributions, which crucially change the high-$T$ stability
of fcc and hcp relative to bcc, is confirmed by our explicit calculation
of these contributions. Further confirmation that fcc is thermodynamically
less stable than bcc at high~$T$ comes from our comparison of the
fcc and bcc melting curves.

At first sight, it might seem unexpected that anharmonicity stabilises
bcc more than fcc and hcp. After all, fcc and hcp are the structures
that go harmonically unstable at $P < 350$~GPa, and intuition
might suggest that below this pressure there could be large,
anharmonically stabilised vibrations, which would have a large entropy.
However, we have seen that the fcc structure at high~$T$ is not
vibrationally unstable, at least with the sizes of simulation cell
that we have used, so presumably phonons that would be harmonically
unstable are stiffened by anharmonic effects, so that their entropy is
actually reduced. In fact, Asker {\em et al.} have shown recently
that electronic thermal excitations have the effect of increasing 
the phonon frequencies of fcc Mo 
(see Fig.~2 in Ref.~[\onlinecite{belonoshko08b}]).      
Furthermore, electronic thermal excitations appear also to further 
stabilize the bcc structure over fcc. 
As we know from previous work, the general effect of high~$T$  
is to smooth the peaks and valleys of the zero-temperature electronic DOS.  
However, in the bcc structure the population of electronic states
on the region near the Fermi energy is enhanced while in the
fcc structure it is depleted. This has the overall effect of  
enhancing the electronic entropy of the bcc structure with respect to 
that of fcc. 
A similar argument has already been suggested by Asker 
{\em et al.}~\cite{belonoshko08b} for explaining the stability properties
of Mo at low~$P$ and high~$T$.   
It is also worth mentioning that in a recent study where we have developed 
a tight-binding model for Mo based on DFT data and used 
it to calculate anharmonic free energies over wide $P - T$ intervals, 
no stabilization of the fcc structure over bcc is observed.~\cite{cazorla09}
The effect of anharmonicity on the thermodynamic
functions of the closely analogous element W has been discussed
recently by Ozolins.~\cite{ozolins09}

Our finding that the fcc melting curve is below the bcc curve,
supported by our own Z-method calculations, is not consistent with the
Mo melting curves deduced by Belonoshko {\em et
  al.}~\cite{belonoshko08} from their Z-method work. We have noted
that one of the difficulties faced by the Z method concerns time
scales. When the temperature $T_{\rm sol}$ of the initially
thermalised crystal exceeds the superheating limit $T_{\rm LS}$, then
in constant-energy MD the system will eventually melt, but the waiting
time $\tau_{\rm w}$ before this occurs may be tens of ps or even more
if $T_{\rm sol}$ is near $T_{\rm LS}$, so that long simulations are
needed if the method is to be reliable.~\cite{Z_long} The time-scale
problem appears to become worse for small systems. The evidence we
have presented indicates that the simulation times of only $\sim 3$~ps
used in the earlier Z-method work~\cite{belonoshko08} were too short
to yield reliable results. The longer simulations of at least $12$~ps
that we use here should give better results, but even so the bcc
melting temperature that we obtain is a significant overestimate
compared with the values from our reference-coexistence
calculations. It would clearly be desirable to repeat the Z-method
calculations with still longer runs on larger systems.  However, the
present simulations do serve the useful purpose of showing that the
Z-method predictions for the relative melting temperatures of bcc and
fcc Mo can be consistent with our much more extensive and detailed
results from free-energy calculations.

The very recent DAC work of Dewaele {\em et al.}~\cite{dewaele10}
on the melting of Ta makes it clear that very careful attention
must be paid to experimental procedures if reliable results are
to be obtained for high-$P$/high-$T$ phase boundaries, and we believe
that a cautious attitude should be adopted towards the existing DAC
evidence~\cite{errandonea01,santamaria09} for low 
melting curves in high-$P$ Mo. Nevertheless,
the shock data on Mo seem to require a crystallographic
boundary somewhere in the region where quasiharmonic calculations
indicate a transition from bcc to fcc or hcp. When we began the
present work, we did not expect that the inclusion of
anharmonicity would cause the bcc-fcc and bcc-hcp boundaries to
disappear. Because we were initially sceptical of our findings,
we felt it essential to confirm them in the ways that we have
described. Our current belief is that efforts should be continued
to search for other candidate crystal structures which might
be thermodynamically more stable than bcc in the high-$P$/high-$T$ region.

In conclusion, our results suggest that the high-$P$/high$T$
solid phase of Mo indicated by shock experiments is not fcc or
hcp, but we do not rule out the possibility of other stable high-$T$
crystal phases. Our results also suggest that the use of
the quasiharmonic approximation should not be uncritically accepted
in the first-principles search for other candidate crystal structures.

\acknowledgments
The work was supported by EPSRC-GB Grant No. EP/C534360, which
was 50\% funded by DSTL(MOD). The work was conducted as
part of a EURYI scheme award to DA as provided by
EPSRC-GB (see www.esf.org/euryi).
The authors acknowledge an allocation time on the HECToR Supercomputer
UK Facilities, some of which was provided by the UKCP consortium.
We also benefited from the UCL Legion High Performance Computing Facilities
and associated support services in the completion of this work.
The authors thank Prof. G. Kresse for help with PAW and Prof. C. J. Pickard
for helpful discussions.


\begin{thebibliography}{99}
\bibitem{paw_radius} We have also checked the effect on the
fcc-bcc free energy difference when we reduce the PAW
core radius. We reduced the core radius from $2.7$~a.u. to $1.6$~a.u.
and we used first-order perturbation theory to estimate the
changes of free energies of fcc and bcc at the state
$V = 8.26$~\AA$^3$/atom and $T = 9000$~K. We found that these free energy
changes amount to only $\sim 10$~meV/atom, which is much smaller
than our computed free energy differences between fcc and bcc.
\bibitem{Z_long} This point was emphasised by the originators of
the Z-method, noting in Ref.~\onlinecite{belonoshko06} that
``...in this method both a rather large number of atoms and long
runs are needed...''.
\end{thebibliography}
\end{document}